\documentclass[a4paper,11pt]{article}

\usepackage[top=1.0in,right=1.0in,left=1.0in,bottom=1.75in]{geometry}
\usepackage{latexsym}
\usepackage{amssymb}
\usepackage{amsmath,amsfonts,amssymb,amsthm}
\usepackage{float}
\usepackage{graphics,graphicx}
\usepackage{tabularx}
\usepackage{amsfonts}
\usepackage{amsmath}
\usepackage{enumerate}
\usepackage{booktabs}
\usepackage{multirow}
\usepackage{sectsty}
\usepackage[affil-it,auth-sc]{authblk}
\usepackage{subcaption}
\usepackage[usenames,dvipsnames]{xcolor}

\usepackage{tabu}

\sectionfont{\normalsize}
\subsectionfont{\normalsize}

\usepackage{fancyhdr}            
\fancypagestyle{plain}{%
  \fancyhead{}                                     
  \fancyfoot[CE,CO]{\thepage}       
  \fancyhead[CE,CO]{\textcolor[rgb]{0.64,0.15,0.15}{Published in \emph{International Journal of Solids and Structures} (2015) \textbf{63} 219-225
\\
doi:10.1016/j.ijsolstr.2015.03.001}}
\setlength{\headheight}{14.5pt}}

\begin{document}

\newcommand{\singlespace}{\baselineskip=12pt\lineskiplimit=0pt\lineskip=0pt}
\def\ds{\displaystyle}

\newcommand{\beq}{\begin{equation}}
\newcommand{\eeq}{\end{equation}}
\newcommand{\lb}{\label}
\newcommand{\ph}{\phantom}
\newcommand{\beqar}{\begin{eqnarray}}
\newcommand{\eeqar}{\end{eqnarray}}
\newcommand{\barr}{\begin{array}}
\newcommand{\earr}{\end{array}}
\newcommand{\jump}{\parallel}
\newcommand{\Ehat}{\hat{E}}
\newcommand{\That}{\hat{\bf T}}
\newcommand{\Ahat}{\hat{A}}
\newcommand{\chat}{\hat{c}}
\newcommand{\shat}{\hat{s}}
\newcommand{\khat}{\hat{k}}
\newcommand{\muhat}{\hat{\mu}}
\newcommand{\mc}{M^{\scriptscriptstyle C}}
\newcommand{\mei}{M^{\scriptscriptstyle M,EI}}
\newcommand{\mec}{M^{\scriptscriptstyle M,EC}}
\newcommand{\hbeta}{{\hat{\beta}}}
\newcommand{\rec}[2]{\left( #1 #2 \ds{\frac{1}{#1}}\right)}
\newcommand{\rep}[2]{\left( {#1}^2 #2 \ds{\frac{1}{{#1}^2}}\right)}
\newcommand{\derp}[2]{\ds{\frac {\partial #1}{\partial #2}}}
\newcommand{\derpn}[3]{\ds{\frac {\partial^{#3}#1}{\partial #2^{#3}}}}
\newcommand{\dert}[2]{\ds{\frac {d #1}{d #2}}}
\newcommand{\dertn}[3]{\ds{\frac {d^{#3} #1}{d #2^{#3}}}}
\newcommand{\ct}{\captionof{table}}
\newcommand{\cf}{\captionof{figure}}
\newcommand{\dd}{\diff}
\newcommand{\rr}{\textcolor{red}}

\def\c{{\circ}}
\def\bob{{\, \underline{\overline{\otimes}} \,}}
\def\ob{{\, \underline{\otimes} \,}}
\def\scalp{\mbox{\boldmath$\, \cdot \, $}}
\def\gdp{\makebox{\raisebox{-.215ex}{$\Box$}\hspace{-.778em}$\times$}}
\def\daa{\makebox{\raisebox{-.050ex}{$-$}\hspace{-.550em}$: ~$}}
\def\mK{\mbox{${\mathcal{K}}$}}
\def\cK{\mbox{${\mathbb {K}}$}}

\def\Xint#1{\mathchoice
   {\XXint\displaystyle\textstyle{#1}}%
   {\XXint\textstyle\scriptstyle{#1}}%
   {\XXint\scriptstyle\scriptscriptstyle{#1}}%
   {\XXint\scriptscriptstyle\scriptscriptstyle{#1}}%
   \!\int}
\def\XXint#1#2#3{{\setbox0=\hbox{$#1{#2#3}{\int}$}
     \vcenter{\hbox{$#2#3$}}\kern-.5\wd0}}
\def\ddashint{\Xint=}
\def\fpint{\Xint=}
\def\dashint{\Xint-}
\def\cpvint{\Xint-}
\def\intl{\int\limits}
\def\cpvintl{\cpvint\limits}
\def\fpintl{\fpint\limits}
\def\ointl{\oint\limits}
\def\bA{{\bf A}}
\def\ba{{\bf a}}
\def\bB{{\bf B}}
\def\bb{{\bf b}}
\def\bc{{\bf c}}
\def\bC{{\bf C}}
\def\bD{{\bf D}}
\def\bE{{\bf E}}
\def\be{{\bf e}}
\def\bbf{{\bf f}}
\def\bF{{\bf F}}
\def\bG{{\bf G}}
\def\bg{{\bf g}}
\def\bi{{\bf i}}
\def\bH{{\bf H}}
\def\bK{{\bf K}}
\def\bL{{\bf L}}
\def\bM{{\bf M}}
\def\bN{{\bf N}}
\def\bn{{\bf n}}
\def\bm{{\bf m}}
\def\b0{{\bf 0}}
\def\bo{{\bf o}}
\def\bX{{\bf X}}
\def\bx{{\bf x}}
\def\bP{{\bf P}}
\def\bp{{\bf p}}
\def\bQ{{\bf Q}}
\def\bq{{\bf q}}
\def\bR{{\bf R}}
\def\bS{{\bf S}}
\def\bs{{\bf s}}
\def\bT{{\bf T}}
\def\bt{{\bf t}}
\def\bU{{\bf U}}
\def\bu{{\bf u}}
\def\bv{{\bf v}}
\def\bw{{\bf w}}
\def\bW{{\bf W}}
\def\by{{\bf y}}
\def\bz{{\bf z}}
\def\T{{\bf T}}
\def\Te{\textrm{T}}
\def\Id{{\bf I}}
\def\bxi{\mbox{\boldmath${\xi}$}}
\def\balpha{\mbox{\boldmath${\alpha}$}}
\def\bbeta{\mbox{\boldmath${\beta}$}}
\def\bepsilon{\mbox{\boldmath${\epsilon}$}}
\def\bvarepsilon{\mbox{\boldmath${\varepsilon}$}}
\def\bomega{\mbox{\boldmath${\omega}$}}
\def\bphi{\mbox{\boldmath${\phi}$}}
\def\bsigma{\mbox{\boldmath${\sigma}$}}
\def\bfeta{\mbox{\boldmath${\eta}$}}
\def\bDelta{\mbox{\boldmath${\Delta}$}}
\def\btau{\mbox{\boldmath $\tau$}}
\def\tr{{\rm tr}}
\def\dev{{\rm dev}}
\def\div{{\rm div}}
\def\Div{{\rm Div}}
\def\Grad{{\rm Grad}}
\def\grad{{\rm grad}}
\def\Lin{{\rm Lin}}
\def\Sym{{\rm Sym}}
\def\Skw{{\rm Skew}}
\def\abs{{\rm abs}}
\def\Re{{\rm Re}}
\def\Im{{\rm Im}}
\def\capB{\mbox{\boldmath${\mathsf B}$}}
\def\capC{\mbox{\boldmath${\mathsf C}$}}
\def\capD{\mbox{\boldmath${\mathsf D}$}}
\def\capE{\mbox{\boldmath${\mathsf E}$}}
\def\capG{\mbox{\boldmath${\mathsf G}$}}
\def\tcapG{\tilde{\capG}}
\def\capH{\mbox{\boldmath${\mathsf H}$}}
\def\capK{\mbox{\boldmath${\mathsf K}$}}
\def\capL{\mbox{\boldmath${\mathsf L}$}}
\def\capM{\mbox{\boldmath${\mathsf M}$}}
\def\capR{\mbox{\boldmath${\mathsf R}$}}
\def\capW{\mbox{\boldmath${\mathsf W}$}}

\def\i{\mbox{${\mathrm i}$}}
\def\mC{\mbox{\boldmath${\mathcal C}$}}
\def\mB{\mbox{${\mathcal B}$}}
\def\mE{\mbox{${\mathcal{E}}$}}
\def\mL{\mbox{${\mathcal{L}}$}}
\def\mK{\mbox{${\mathcal{K}}$}}
\def\mV{\mbox{${\mathcal{V}}$}}
\def\C{\mbox{\boldmath${\mathcal C}$}}
\def\E{\mbox{\boldmath${\mathcal E}$}}

\def\AAM{{\it Advances in Applied Mechanics }}
\def\ACME{{\it Arch. Comput. Meth. Engng.}}
\def\ARMA{{\it Arch. Rat. Mech. Analysis}}
\def\AMR{{\it Appl. Mech. Rev.}}
\def\ASCEEM{{\it ASCE J. Eng. Mech.}}
\def\ACTA{{\it Acta Mater.}}
\def\CMAME {{\it Comput. Meth. Appl. Mech. Engrg.}}
\def\CRAS{{\it C. R. Acad. Sci. Paris}}
\def\CRM{{\it Comptes Rendus M\'ecanique}}
\def\EFM{{\it Eng. Fracture Mechanics}}
\def\EJMA{{\it Eur.~J.~Mechanics-A/Solids}}
\def\IJES{{\it Int. J. Eng. Sci.}}
\def\IJF{{\it Int. J. Fracture}}
\def\IJMS{{\it Int. J. Mech. Sci.}}
\def\IJNAMG{{\it Int. J. Numer. Anal. Meth. Geomech.}}
\def\IJP{{\it Int. J. Plasticity}}
\def\IJSS{{\it Int. J. Solids Structures}}
\def\IngA{{\it Ing. Archiv}}
\def\JAM{{\it J. Appl. Mech.}}
\def\JAP{{\it J. Appl. Phys.}}
\def\JAE{{\it J. Aerospace Eng.}}
\def\JE{{\it J. Elasticity}}
\def\JM{{\it J. de M\'ecanique}}
\def\JMPS{{\it J. Mech. Phys. Solids}}
\def\JSV{{\it J. Sound and Vibration}}
\def\MACRO{{\it Macromolecules}}
\def\MMT{{\it Mech. Mach. Th.}}
\def\MOM{{\it Mech. Materials}}
\def\MMS{{\it Math. Mech. Solids}}
\def\MMT{{\it Metall. Mater. Trans. A}}
\def\MPCPS{{\it Math. Proc. Camb. Phil. Soc.}}
\def\MSE{{\it Mater. Sci. Eng.}}
\def\NATURE{{\it Nature}}
\def\NATUREM{{\it Nature Mater.}}
\def\PHIL{{\it Phil. Trans. R. Soc.}}
\def\PMPS{{\it Proc. Math. Phys. Soc.}}
\def\PNAS{{\it Proc. Nat. Acad. Sci.}}
\def\PRE{{\it Phys. Rev. E}}
\def\PRL{{\it Phys. Rev. Letters}}
\def\PRSL{{\it Proc. R. Soc.}}
\def\RIIT{{\it Rozprawy Inzynierskie - Engineering Transactions}}
\def\ROCK{{\it Rock Mech. and Rock Eng.}}
\def\QAM{{\it Quart. Appl. Math.}}
\def\QJMAM{{\it Quart. J. Mech. Appl. Math.}}
\def\SCIENCE{{\it Science}}
\def\SCRMAT{{\it Scripta Mater.}}
\def\SM{{\it Scripta Metall.}}
\def\ZAMM{{\it Z. Angew. Math. Mech.}}
\def\ZAMP{{\it Z. Angew. Math. Phys.}}
\def\ZVDI{{\it Z. Verein. Deut. Ing.}}

\renewcommand\Affilfont{\itshape\small}
\setlength{\affilsep}{1em}
\renewcommand\Authsep{, }
\renewcommand\Authand{ and }
\renewcommand\Authands{ and }
\setcounter{Maxaffil}{3}

\renewcommand{\vec}[1]{{\boldsymbol{#1}}}

\definecolor{traz}{RGB}{200,50,0}
\definecolor{comp}{RGB}{0,30,200}

\newcommand{\djc}[1]{{\color{Orange}#1}}

\title{Experimental and analytical insights on fracture trajectories in brittle materials with voids}

\author{D. Misseroni$^1$,  A.B. Movchan$^1$,  N.V. Movchan$^1$ and D. Bigoni$^2$\footnote{Corresponding author:\,e-mail:\,bigoni@ing.unitn.it; phone:\,+39\,0461\,282507.}\\
\normalsize{e-mail: diego.misseroni@ing.unitn.it, abm@liverpool.ac.uk;}\\
\normalsize{ nvm@liverpool.ac.uk, bigoni@ing.unitn.it;}\\
\small{$^1$ University of Liverpool, Liverpool L69 7ZL,  United Kingdom}\\
\small{$^2$ DICAM - University of Trento, I-38050 Trento, Italy.}
}
\maketitle

\begin{abstract}

Analytical models have been developed for fracture propagation over the last several decades and are now considered with renewed interest;
the range of their applicability varies for different materials and different loading conditions.
Systematic experimental measurements and numerical simulations are presented against an asymptotic prediction for the crack trajectory developing
in solids with small voids (of circular and elliptical shape).
The experiments also cover the dynamic regime, where new features involving crack kinking and roughness of the fracture surface occur.

\end{abstract}


\section{Introduction}

Fracture is the most limiting factor for the development of highly brittle materials such as advanced ceramics.
For these materials, theoretical predictions of the geometrical shape of cracks are rare 
and when such predictions are possible they are always valuable both for design
purposes and for the advances in the knowledge of these fragile media.
The deflection of a crack from a straight line is naturally observed in experiments and linked to an increase of toughness \cite{faber, Mirkhalaf, giovanni}. 
Crack trajectories may be deflected from rectilinearity 
due to an interaction with imperfections such as small voids or inclusions, which may result from exploitation damage or manufacturing defects. Crack branching and  path deflection in two-dimensional elastic solids were examined by Sumi et al. (1983) \cite{Sumi}, where the asymptotic method was also illustrated for the case of a slightly slanted Griffith crack under bi-axial load.
Computational models have been developed by Xu et al. (1994; 1998) \cite{Bower}\cite{Xu} to study the crack growth in heterogeneous solids, while an energy minimizer technique for crack path determination has been developed by
Francfort and Marigo (1998) \cite{Francfort2} and Dal Maso et al. (2005) \cite{Francfort}.
The local higher-order asymptotic analysis for the elastic fields near the crack tip was developed by Sumi (1986)\cite{Sumi3} and the idea of using the higher-order terms to predict a fracture path was implemented in the computational FEM algorithm\cite{Sumi2}: to our knowledge this was the first systematic attempt towards the formulation of the hybrid method, based on superposition of analytical and numerical solutions, employing the coefficients in the first-order perturbation formulae. The paper by Sumi (1986)\cite{Sumi3} presented simulations for both kinking and a smooth crack advance.  The finite element simulation for a system of cracks, based on the first-order perturbation method and an incremental quasi-static crack growth assumption, was developed by Sumi and Wang (1998)\cite{Sumi4}. Formulations for a curved crack based on a perturbation procedure were discussed by Hori and Vaikuntan (1997)\cite{Hori}, although no weight functions were involved in their analysis.

Two-dimensional models (both plane stress and plane strain) of crack propagation have been developed in the analytical form, in which the crack path was predicted in terms of elementary functions or via a straightforward integration involving weight functions.
These approaches
include an asymptotic model describing the
interaction of a semi-infinite crack with small voids \cite{Movchan},\cite{Movchan2},\cite{Movchan3},\cite{Movchan4}, where the voids are
characterized by their dipole tensors \cite{polya}, also known as the virtual mass tensor. In the case of vector elasticity, the dipole tensors are of rank 4, but allow for a matrix representation in a specially chosen basis; the corresponding matrices are referred to as P\'olya-Szeg\"o matrices \cite{Movchan4}.

\begin{figure}[h]
  \begin{center}
      \includegraphics[width= 15cm]{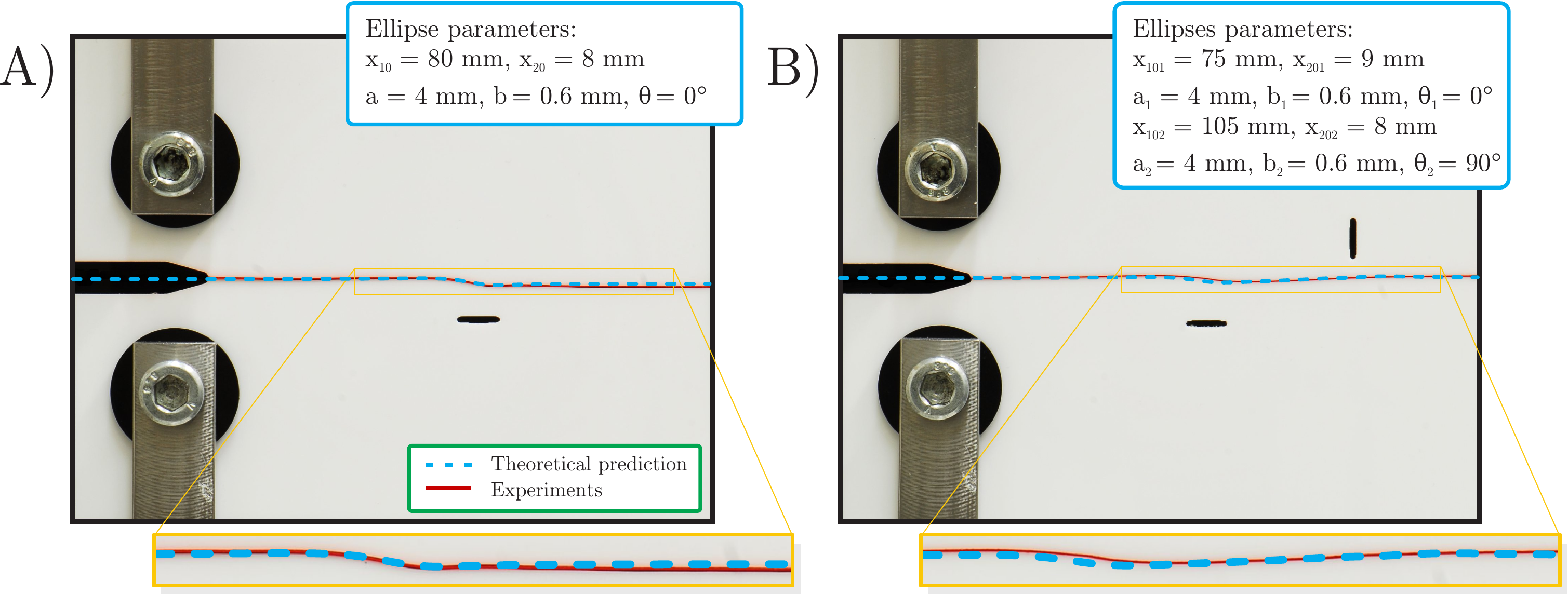}
\caption{\footnotesize {Experimental crack trajectory (a post-mortem photo of a PMMA sample loaded in Mode-I) compared to the theoretical prediction (reported with a cyan/dashed
line) for two cases of interaction with one (A) and two (B) elliptical voids. The crack trajectory is highlighted by means of a solid/red colored strip inserted between the two crack surfaces. The parameters indicated in the figure are explained in Fig. \ref{fig1-mo}.} }
\label{Exp_ellipses2}
  \end{center}
\end{figure}

Practical needs require efficient tools, and the asymptotic model, appealing for its simplicity, can be used for many fracture configurations. In fact,
Bigoni et al. (1998) \cite{Valentini1} and Valentini et al. (1999, 2002) \cite{Valentini2} \cite{Valentini0} have successfully compared theoretical predictions with experimental measurements, but in an indirect way, i.e. the cracks were induced by an indenter, which provided a complex stress state, different from the remote Mode-I condition assumed in the model.
In the present article, the reliability of the asymptotic model is assessed for quasi-static regimes, together with the observation of important effects related to dynamic crack kinking.
In particular, in addition to the asymptotic model,
a computational technique, namely, the
extended finite element method (X-FEM) as implemented in Abaqus FEA (Shi et al., 2010 \cite{Shi2010}), is employed to predict crack propagation.
This method (already used for the analysis of structural components failure \cite{Miranda2002}, \cite{Blittencourt1996} and of quasi-static crack growth \cite{Sukumar2003}, \cite{Belytschko2001}) is employed, together with the asymptotic model, to predict crack paths in brittle (PMMA) notched plates subject to Mode-I loading and containing elliptical voids.
An excellent agreement between theoretical results and experiments is observed, with the asymptotic solution
correctly predicting fracture deflection toward the small void, as shown for instance in
Fig. \ref{Exp_ellipses2}, where a fracture path deviates from the straight line due to the interaction with thin elliptical voids.
In the figure the cyan/dashed  curve represents the shape predicted by the asymptotic model, whereas the solid/red curve shows the actual shape obtained in the experiment (the FE simulation coincides with the asymptotics and hence is not reported in the figure).

From this result it can be deduced that the asymptotic approach fully explains crack deflection induced by a void (when sufficiently far from the fracture trajectory to satisfy the asymptotic
assumptions), so that it can be employed
in design situations involving the presence of defects in ceramics.

\section{Modelling}

\subsection{Asymptotic model}

In this section we give a brief outline of 
the model used for the analysis of crack trajectories influenced by elliptical voids; for a more detailed
description see Movchan et al. (1991)\cite{Movchan2}, Movchan (1992)\cite{Movchan3}, Movchan and Movchan (1995)\cite{Movchan4}, Movchan et al. (1997)\cite{Movchan2}, Valentini et al. (1999)\cite{Valentini2}.

\begin{figure}[h]
  \begin{center}
    \vspace*{5mm}
    \includegraphics[width=15.0cm]{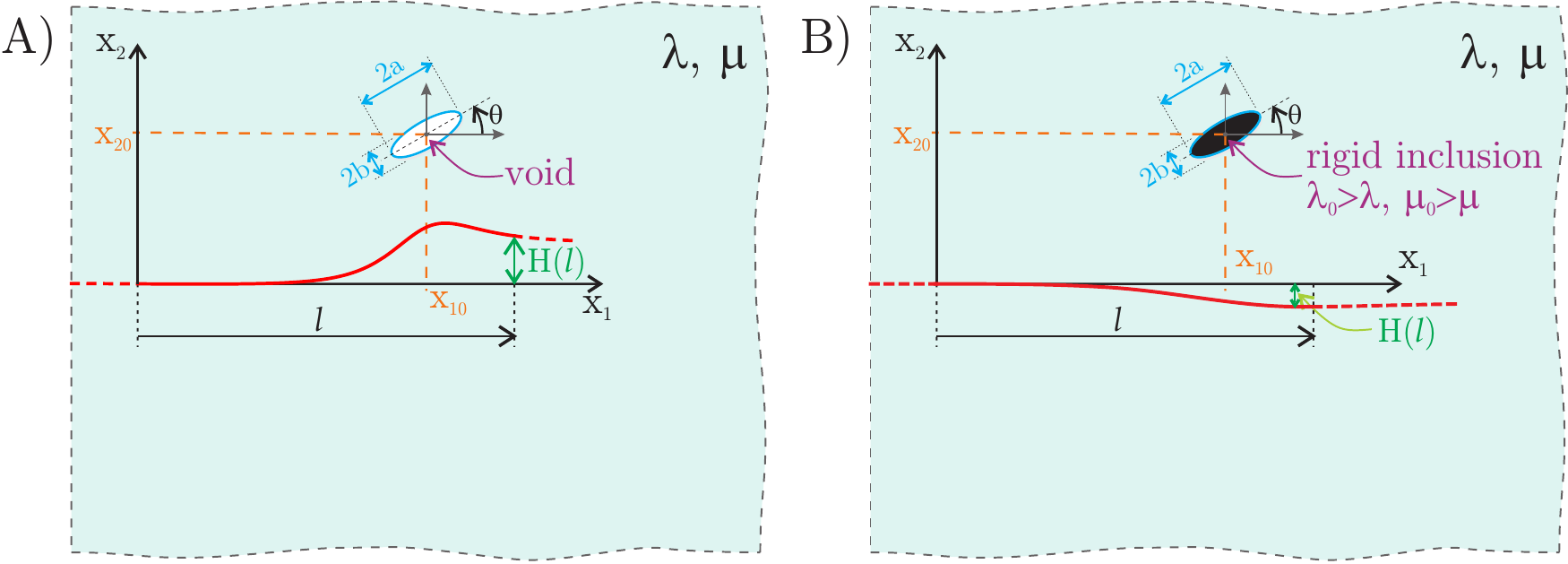}
    \caption{\footnotesize{The analyzed crack geometry interacting with an elliptical void (A) and an elliptical elastic inclusion (B).}}
    \label{fig1-mo}
  \end{center}
\end{figure}

An infinite, brittle isotropic linearly elastic body is assumed, containing a semi-infinite crack interacting with some defects (voids or inclusions), and
propagating under pure Mode-I loading, corresponding to a stress intensity factor $K_I$ greater than the critical one.
The elastic properties are specified through the Lam\'{e} constants $\lambda$ and $\mu$ for the body and $\lambda_{0}$ and $\mu_{0}$ for the inclusion.
We assume that the defects present in the infinite elastic medium are \lq far' from the straight trajectory
that would be followed by the crack in the absence
of disturbances (Fig.\,\ref{fig1-mo}).
The shape of the defect is selected to be an ellipse centered at the point ($x_{1}^{0},x_{2}^{0}$) with the
major and minor semi-axes denoted by $a$ and $b$, respectively.
The major axis is inclined at an angle $\theta$ with respect to the $x_{1}$-axis.
Due to the presence of the defect the crack trajectory may deviate from a straight line and this perturbation  can be described by the closed-form formula for the crack trajectory
$H(l)$, expressed as a function of the crack tip coordinate $l$ (Valentini et al., 1999; 2002)\cite{Valentini2} \cite{Valentini0}
\begin{small}
\begin{align}
  H&(l) = \frac{ab}{4x_{2}^{0}}\bigg\{\Upsilon_0 (\rho^{4}-\Theta_0)(t+t^{2}-2) +2\rho^{2}\Upsilon_0 \Theta_0 \bigg[ \nonumber\\
  &\sin {2\theta }(t+t^{2})(2t-1) \sqrt{1-t^{2}}
  -\cos {2\theta}(t-t^{3})(1+2t)\bigg]  \nonumber\\
  &+\frac{\Theta_0 \rho^{4}}{\rho^{4}+\Theta_0}
  (t-t^{3})\bigg[ \left( 1+\Xi_0 \cos^{2}{2\theta }\right) (1-t)(1+2t)^{2}  \nonumber\\
  &+\left( 1+\Xi_0 \sin^{2}{2\theta }\right) (1+t)(2t-1)^{2}\nonumber\\
  &
  -\Xi_0 \sin {4\theta}\sqrt{1-t^{2}}(4t^{2}-1)\bigg] \bigg\},
  \label{uno}
\end{align}
\end{small}
where the variable $t$ depends on $l$ through the relation
\begin{equation}
 t=\frac{x_{1}^{0}-l}{\sqrt{\left( x_{2}^{0}\right)^{2}
 +\left( x_{1}^{0}-l\right)^{2}}},  \quad  \mbox{and}  \quad
 \rho=\sqrt{\frac{a+b}{a-b}}.
\label{due}
\end{equation}

The elastic properties of the inclusion and matrix influence the crack trajectory via the
constants:
\begin{equation}
\begin{aligned}
  \Theta_0 &= \frac{\mu_{0}-\mu }{\kappa \mu_{0}+\mu },  \\[2mm]
  \Xi_0 &= \frac{2\Theta_0 (\kappa +1)\mu_{0}}{\rho^{4}[(\kappa_{0}-1)\mu +2\mu_{0}]
  +\Theta_0 [(\kappa_{0}-1)\mu -2\kappa \mu_{0}]},  \\[2mm]
  \Upsilon_0 &= \frac{2[(\kappa -1)\mu_{0}-(\kappa_{0}-1)\mu ]}{\rho^{4}
  [(\kappa_{0}-1)\mu +2\mu_{0}]+\Theta_0 [(\kappa_{0}-1)\mu -2\kappa \mu_{0}]},
  \label{tre}
\end{aligned}
\end{equation}
where $\kappa =3 - 4\nu$ (for plane strain) or $\kappa =(3-\nu)/(1+\nu)$ (for plane stress), and $\nu\in(-1,1/2)$ is the Poisson's ratio.

In the  case of \emph{plane stress},  the formula for the crack trajectory due to elliptical void may be
obtained by taking $\mu_0=\lambda_0=0$ in (\ref{uno})-(\ref{tre}), which gives
\begin{multline}
  H(l)=\frac{(1-\nu^2)R^{2}}{2x_{2}^{0}}\bigg[ 2(1+m^{2})-t\bigg( 2+t-t^{2}+m^{2}(1+t) \\
  +2m\cos {2\theta }(1+2t)(1-t^{2})-2m\sin {2\theta }(2t-1)(1+t)\sqrt{1-t^{2}}\bigg)\bigg],
  \label{quattro}
\end{multline}
where
\begin{equation*}
  R=\frac{a+b}{2}   \quad \mbox{and} \quad   m=\frac{a-b}{a+b}.
\end{equation*}

It is important to remark that expression (\ref{quattro}) depends on the Poisson's ratio $\nu$, the angle of inclination $\theta$ of the major
axis, and the parameters $R$ and $m$.

We note that imperfectly bonded circular inclusions have been
considered in Bigoni et al. (1998)\cite{Valentini1}.

\subsection{Adaptive numerical simulation}

An algorithm has been implemented in ABAQUS to simulate a quasi-static crack growth in a solid with small voids.
It enables a crack advance, according to a brittle fracture criterion, without any remeshing.

We have computed the trajectory for crack propagation by means of the 2D extended finite element method (X-FEM), implemented in the Abaqus/Standard finite element package.
The eXtended Finite Element Method is a partition of unity based method, particularly suitable for modelling crack propagation phenomena, without a-priori knowledge of the crack path and without remeshing.
The simulations were performed using a parametric python script for ABAQUS, that is run by means of MatLab. CPS4 elements were used for 2D plane stress analyses.

\begin{figure}[h]
  \begin{center}
      \includegraphics[width= 15 cm]{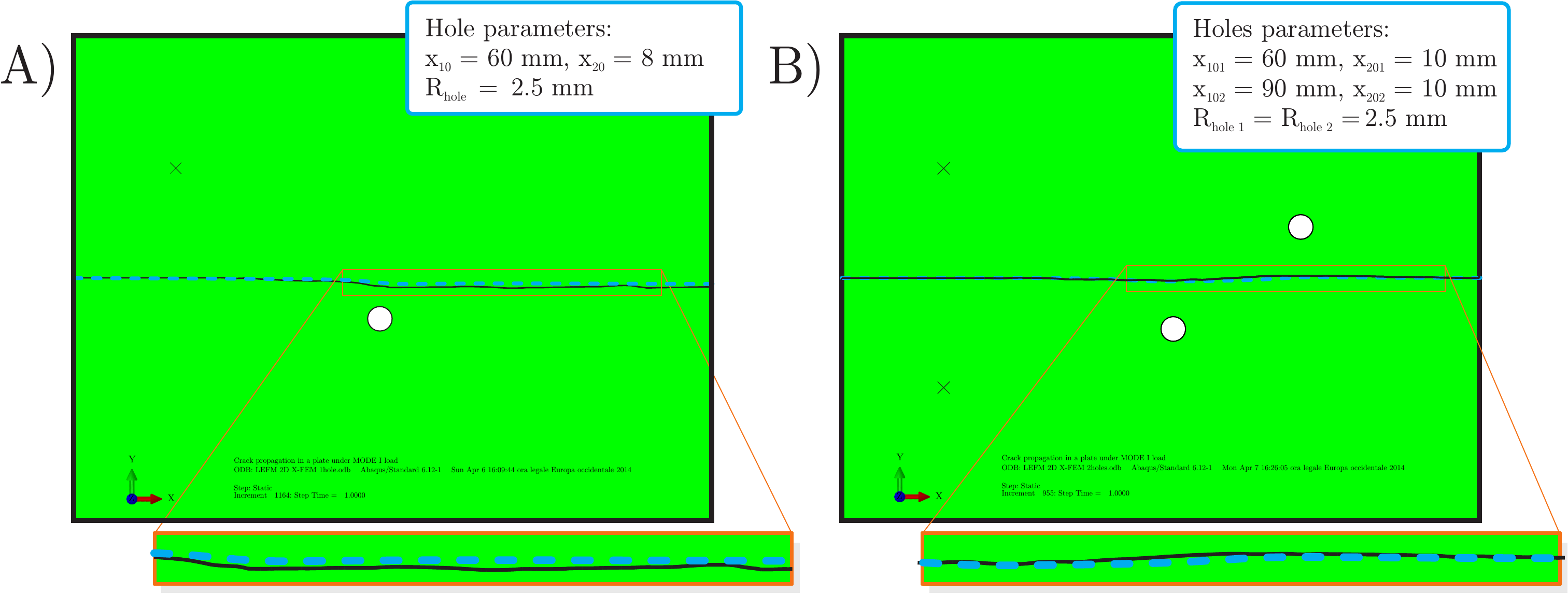}
\caption{\footnotesize {X-FEM Abaqus simulations under the plane stress assumption for one (A) and two (B) circular voids are reported (solid/black line) and compared to results obtained with  the asymptotic formulation (cyan/dashed line). }} \lb{Abaqus}
  \end{center}
\end{figure}

The crack trajectory evolution was modelled via Linear Elastic Fracture Mechanics approach (LEFM) specific for brittle fracture. In particular, the strain energy release rate at the crack tip was calculated using the virtual crack closure technique (VCCT), so that
the crack propagates when the strain energy release rate reaches a critical value. We have chosen the so-called \lq BK law' to simulate mix-mode behavior (tolerance=0.1, viscosity=0.0001) and the Maximum Tangential Stress (MTS) as the criterion for singling out the crack propagation direction. The former criterion is a combination of the energy release rates in Mode I, Mode II, and Mode III in a power law relationship (Benzeggagh and Kenane, 1996) \cite{Benzeggagh}. The latter criterion uses the maximum tangential stress  to determine the normal direction to the crack plane (Erdogan and Sih, 1963) \cite{Erdogan}. 
Note that the X-FEM approach allows us to avoid the use of special finite elements with embedded singularities, which 
requires that the crack propagates across an entire element at a time. 
The strain energy release rate at the crack tip has been computed by the modified Virtual Crack Closure Technique (VCCT) in combination with principles of linear elastic fracture mechanics. The VCCT is based on the assumption that the strain energy released for crack extension of a certain amount is the same as the energy required to close the crack of the same amount.

The computation set-up, in term of sample size, the way to apply the load and the material properties were selected to be identical to the experimental set-up; in particular,
the load is applied by imposing a displacement at the grips of 0.8 $\mu$m/s, in order to maintain a low loading rate.

Two examples of the X-FEM crack path predictions are shown in Fig. \ref{Abaqus} and compared to the prediction of the asymptotic model, eq. (\ref{quattro}), for one and two circular voids. An excellent agreement can be observed.

An asymptotic formula for the crack trajectory interacting with a small void, characterized by a $3 \times 3$ dipole matrix $\mathcal{P}$ is given in \cite{Valentini2}, see also \cite{Movchan2}.  The following formula is valid for a crack interacting with an arbitrary small void (not necessarily elliptical)

\begin{equation}
  H(l)= \frac{4 \mu}{x_{2}^{0}(\kappa+1)}\big[\cos \phi  \mathcal{L}^t(\phi)\mathcal{P}\mathcal{L}(\phi)-\mathcal{L}^t(0)\mathcal{P}\mathcal{L}(0)
 \big],
  \label{dipole_expr}
\end{equation}
with $\cos \phi=(x_1^0-l)\big[(x_2^0)^2+(x_1^0-l)^2\big]^{-1/2}$ and
the vector function $\mathcal{L}$ defined by

\begin{equation}
\mathcal{L}(\phi, \kappa)=
\begin{pmatrix}
\ds \frac{1}{4 \mu \sqrt{2 \pi}} \cos \frac{\phi}{2} \big[ \kappa -1-2 \sin \frac{\phi}{2}\sin\frac{3\phi}{2} \big]\\
\ds \frac{1}{4 \mu \sqrt{2 \pi}} \cos \frac{\phi}{2} \big[ \kappa -1+2 \sin \frac{\phi}{2}\sin\frac{3\phi}{2} \big] \\
\ds \frac{1}{4 \mu \sqrt{2 \pi}} \sin \phi \cos \frac{3 \phi}{2}
\end{pmatrix}.
\end{equation}

In the specific case of an elliptical void, the dipole matrix $\mathcal{P}$ becomes (Movchan et al. (1997)\cite{Movchan2})
\begin{align}
\mathcal{P}(\kappa)= \frac{1}{4\,q}
\begin{pmatrix}\scriptstyle
\ds \frac{(\kappa - 1)[\Sigma- R^2(\kappa - 1)]-\Xi}{(\kappa - 1)^2} & R^2 - \ds \frac{\Xi}{(\kappa - 1)^2} & \ds \frac{\Lambda}{\kappa-1}\\
R^2 - \ds \frac{\Xi}{(\kappa - 1)^2}  & \ds \frac{(\kappa - 1)[-\Sigma- R^2(\kappa - 1)]-\Xi}{(\kappa - 1)^2} &  \ds \frac{\Lambda}{\kappa-1}\\
 \ds \frac{\Lambda}{\kappa-1} &  \ds \frac{\Lambda}{\kappa-1} & -2 R^2
\end{pmatrix},
\end{align}
where
\begin{equation}
\begin{aligned}
q=\frac{\lambda+\mu}{8\pi\mu(\lambda+2\mu)}, \hspace{2mm}
  \Sigma = 4 m R^2 \cos 2\theta,  \\
  \Xi = 2 R^2 (1+m),  \hspace{2mm}  \Lambda = 2 \sqrt{2} m R^2 \sin 2\theta.
    \label{costanti}
\end{aligned}
\end{equation}

The crack deflection at infinity, \lq far away' from the small void, can be obtained from eq. (\ref{dipole_expr}) in the limit $l \rightarrow  \infty$, which for a generic void becomes
\begin{equation}
  H(\infty)=\frac{4\mu}{\kappa + 1}[\mathcal{L}^t(0)\mathcal{P}\mathcal{L}(0)] ,
  \label{value_infinity}
\end{equation}
and in the particular case of an ellipse
\begin{equation}
  H(\infty)=\frac{R^{2}}{x_{2}^{0}}(1+m^{2}).
  \label{value_infinity}
\end{equation}

We would like to emphasize that, although the deflection at infinity depends on the dipole matrix $\mathcal{P}$, it is independent of the void's orientation (independent of $\theta$ for the elliptical void).
This important feature is clearly visible from both the Abaqus simulations and the experiments (see Figs. \ref{Exp_ellipses2}, \ref{Abaqus}, \ref{Exp_holes} and \ref{Exp_ellipses_1}).

\section{Experiments}

The experimental verification of the theoretical asymptotic model outlined above was performed (at the \lq Instabilities Lab', http://www.ing.unitn.it/dims/ssmg/)  by quasi-statically loading (controlled displacements) notched plates of a brittle material under Mode-I conditions until failure. The post-mortem crack trajectories were analyzed and compared to the predictions of the asymptotic method and 
to those of the numerical model. 
We highlight the limitations of the asymptotic approach to the cases of quasi-static crack advance, showing that in the dynamic regime an instability may occur, leading to a crack kinking. The configurations reported in the text include several geometries of small voids  at different orientations. The implementation of the experiment was done to ensure the required quasi-static growth of a crack.

\subsection{Quasi-static crack propagation}

PMMA (PERPEX white and black from Bayer, with elastic modulus 3350 MPa,
$\nu$=0.38) plates with a V-shaped notch (to be loaded in-plane) were employed for the study of crack propagation,
in agreement with the standard test method ASTM E647-00.
Samples of two different dimensions were used, namely, 125 mm x 95.0 mm x 3.0 mm and 130 mm x 105 mm x 3mm, for testing one or more circular or elliptical voids, respectively.
The notches have been engraved with a cutter (realized from a blade of a Stanley 99E utility knife to advance of a fixed amount of 0.4 mm with a micrometrical screw, see Fig. \ref{Exp_holes} C) to trigger the propagation of the fracture.
Specimens were obtained by cutting a PMMA plate with a EGX-600 Engraving Machine (by Roland).
The position and the size of the voids are reported in the labels of the figures, with reference to the nomenclature introduced in Fig. \ref{fig1-mo} A.

Tensile force on the structure was provided by imposing displacement with a load frame MIDI 10 (Messphysik) and the load measured
with a load cell Gefran TH-KN2D. Data was acquired with a NI CompactDAQ system, interfaced with Labview 8.5.1 (National Instruments). Photos
were taken with a Sony NEX 5N digital camera (equipped with 3.5-5.6/18-55 lens, optical
steady shot from Sony Corporation), with a Nikon D200 digital camera (equipped with a AF-S micro Nikkor lens 105 mm 1:2.8G ED), or with a stereoscopic microscope Nikon SMZ-800 (equipped with objectives P-Plan Apo 0.5X and P-ED Plan 1.5X).
A very low rate of loading, namely, 0.8 $\mu$m/s was imposed during the test to ensure quasi-static crack propagation.

\subsection{Experimental measurements versus analytical prediction}

For the samples where small (circular and elliptical) voids lie at different relative positions and orientations, but always sufficiently distant, from the propagating crack, we demonstrate an impressive agreement between the theoretical prediction of the asymptotic model and the experimental measurements. It is also noted that the crack deflection at infinity is approximately the same for different  orientations of the voids of non-circular geometry, which is in agreement with the prediction of the asymptotic model.

\subsubsection{Circular versus elliptical void attracting the crack }

Experiments for both circular and elliptical voids, were performed, Figs. \ref{Exp_ellipses2}, \ref{Exp_holes}, \ref{Exp_ellipses_1}, and compared to theoretical results obtained with the asymptotic approach, eq. (\ref{quattro}).

The numerical simulations were found to match the the predictions of the asymtotic so well that they were superimposed on the predictions of the asymptotic model and therefore have not been reported.

\begin{figure}[h]
  \begin{center}
      \includegraphics[width= 15 cm]{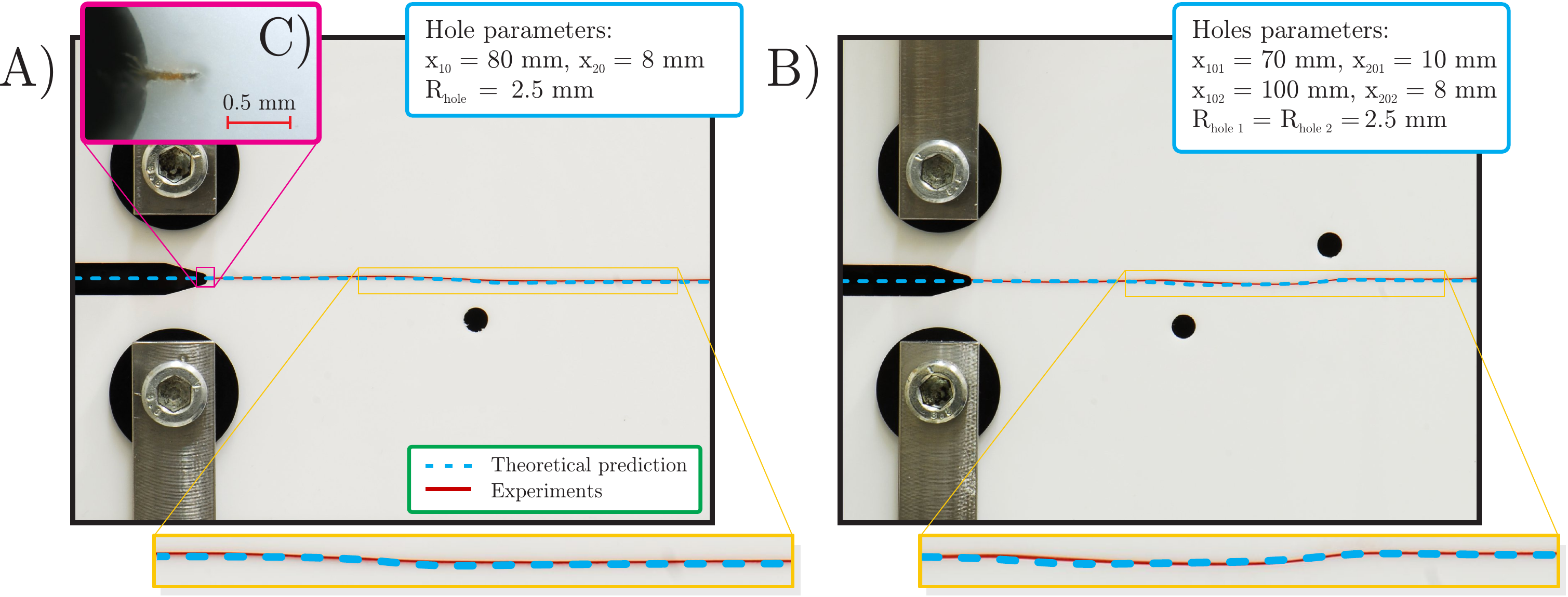}
\caption{\footnotesize {Experimental crack trajectory (a post-mortem photo of a PMMA sample loaded in Mode-I) compared to the theoretical prediction (reported with a cyan/dashed
line) for two cases of interaction with one (A) and two (B) circular voids. The crack trajectory is highlighted by means of a solid/red colored strip inserted between the two crack surfaces.
(C) Detail of the engrave cut at the notch tip to trigger rectilinear fracture propagation.}
} \lb{Exp_holes}
  \end{center}
\end{figure}

\begin{figure}[h]
  \begin{center}
      \includegraphics[width= 15 cm]{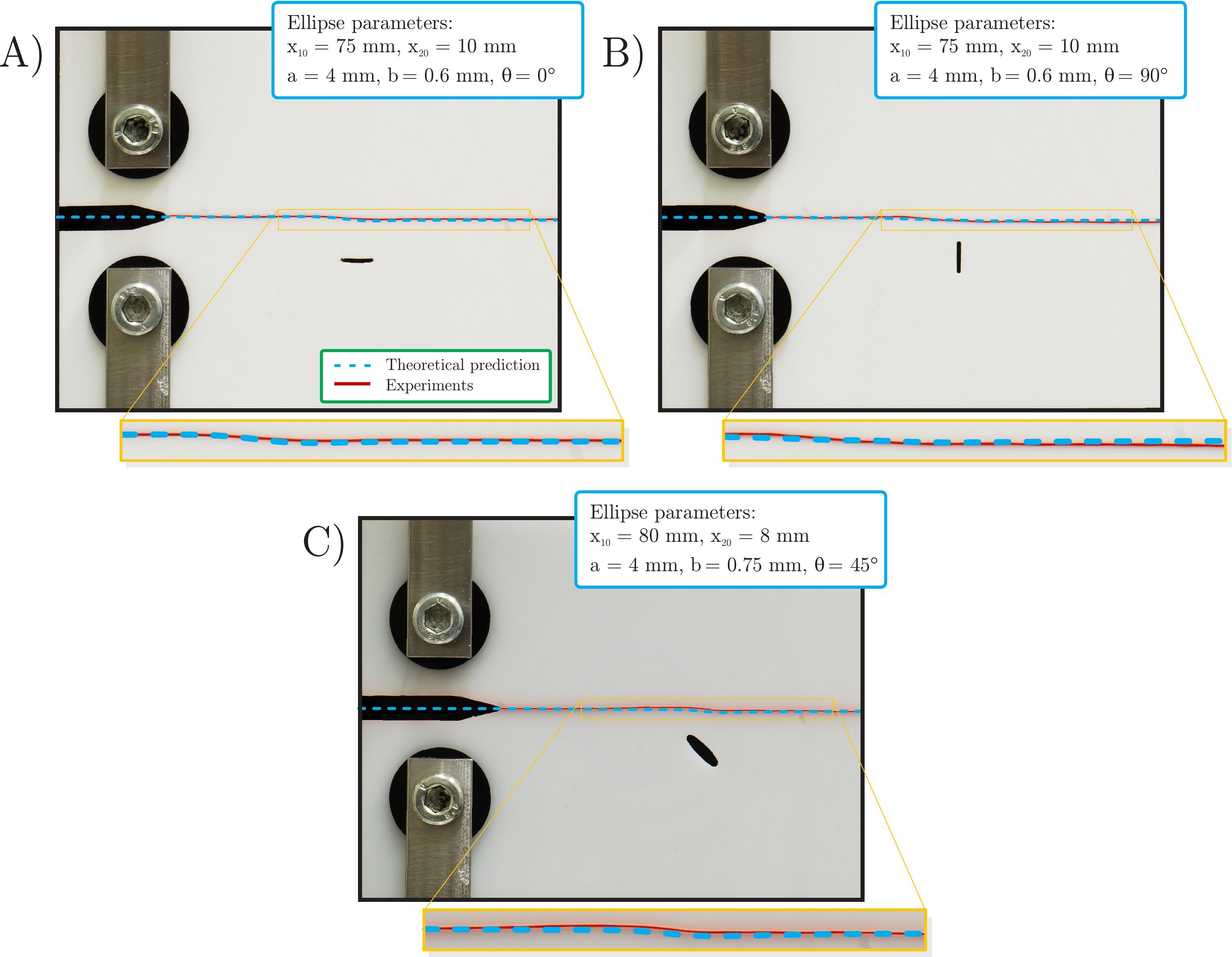}
\caption{\footnotesize {Experimental crack trajectory (a post-mortem photo of a PMMA sample loaded in Mode-I) compared to the theoretical prediction (reported with a white cyan/dashed
line) for two cases of interaction with one elliptical void: major axis is inclined at an angle $\theta=0^\circ$ (A), $\theta=90^\circ$ (B) and $\theta=45^\circ$ (C) with respect to the $x_{1}$-axis. The crack trajectory is highlighted by means of a solid/red colored strip inserted between the two crack surfaces.}} \lb{Exp_ellipses_1}
  \end{center}
\end{figure}

Two experiments for one and two voids are reported in each of Figs. \ref{Exp_ellipses2} and \ref{Exp_holes}; elliptical voids were considered in the former figure and circular in the latter.
Beside the very satisfactory theoretical/experimental agreement, it is important to remark that, in the case of two voids (elliptical or circular), the deflection at infinity is close to zero, as a consequence of the sum of two trajectory deflections with opposite sign.

The influence of the orientation of an elliptical void has been investigated in Fig. \ref{Exp_ellipses_1}, where three experiments are reported, all confirming the theoretical prediction that the inclination of the ellipses does not influence crack deflection at infinity.

In conclusion, the two most important phenomena that are very evident from the experiments are that the voids \lq attract' the crack trajectory and that the deflection at infinity only depends on the dipole matrix as predicted from the asymptotic model, eq. (\ref{value_infinity}).


\subsection{Applicability of the asymptotic model, dynamics and crack kinking}

In this section we highlight the limitations of the above-described asymptotic model in the case when the crack trajectory is too close to the void and when propagation of the crack reaches the dynamic regime.

The case of a large circular void close to the rectilinear crack trajectory (that would occur in the absence of the void) was investigated and results are reported in Fig. \ref{palla}, for the experimental fracture trajectory (Fig. \ref{palla} A)  and its asymptotic and numerical predictions (Fig. \ref{palla} B).
\begin{figure}[h]
  \begin{center}
      \includegraphics[width= 15 cm]{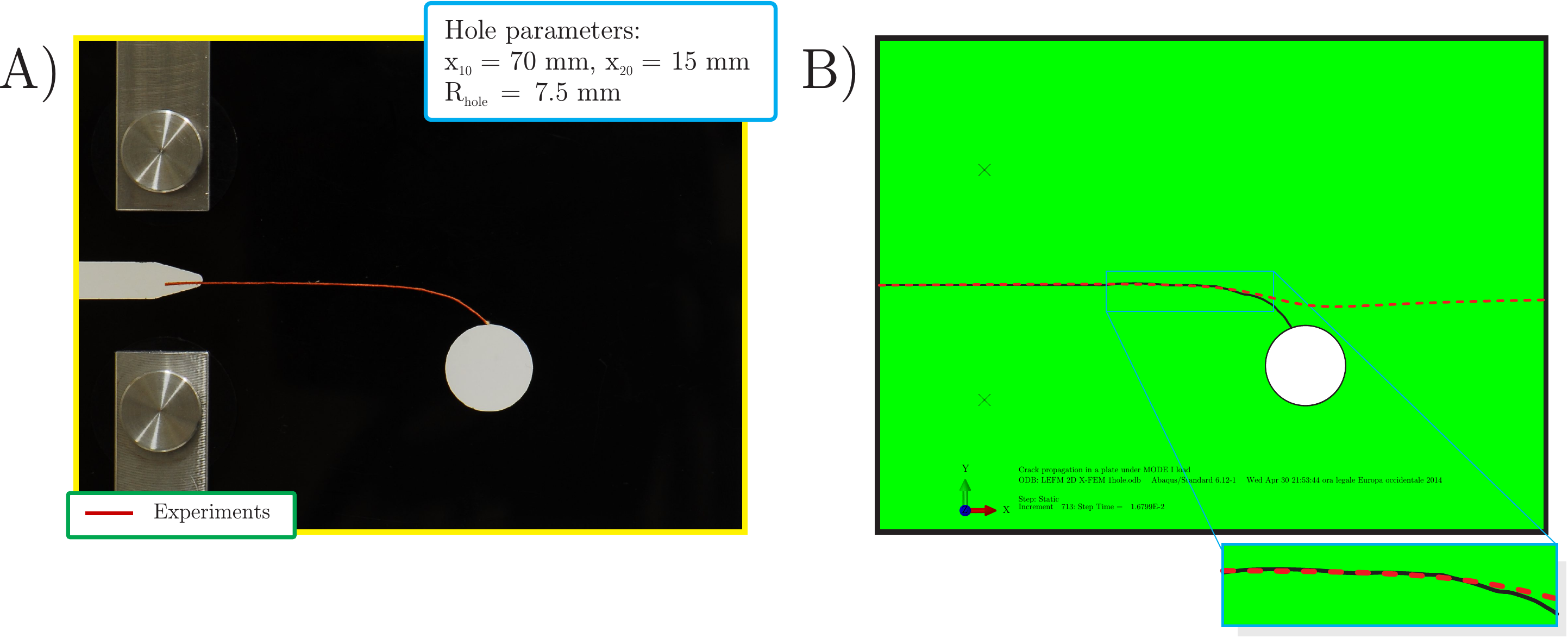}
\caption{\footnotesize { Comparison between experimental (A) and theoretical (B) results for quasi-static crack trajectories in the case of high value of the ratio between the void diameter and the distance from the void centre to the crack trajectory. The numerical simulation correctly predicts the crack path, while the asymptotic approach correctly captures only the initial crack deflection.}} \lb{palla}
  \end{center}
\end{figure}
The asymptotic model is based on the assumptions that
the ratio $\epsilon$ between the inclusion diameter and the distance from the inclusion centre to the crack trajectory is small, an assumption now clearly violated.
It is interesting to notice that the crack trajectory can be captured with the numerical model, while the asymptotic evaluation still provides a good approximation to the {\it initial} crack deviation.

The other departure from the hypotheses of both the asymptotic and the numerical approach occurs when
a sufficiently high (and constant) rate of loading is applied to the sample. In this case, an acceleration of the crack occurs, leading subsequently to a dynamic kinking of the crack trajectory.
The latter differs from the case of the quasi-static crack propagation, where the crack path remains smooth; the loading rate has to be reduced significantly (as done in the previously presented experiments)
in order to achieve conditions appropriate for a quasi-static crack growth.
From the experiments we have seen that there is a wide range of applied displacement rate within which the asymptotic model gives a good prediction of the quasi-static crack trajectory.
In particular, if the loading rate determined by the speed of the loading movable crosshead (shown in Fig. \ref{Dynamic_VS_Static}) is constant and exceeds 10 mm/s, then the crack growth accelerates and becomes unstable, resulting in a roughness of the fracture surface and kinking of the crack trajectory. On the other hand, if the loading rate falls below 10 mm/s, the crack path remains smooth and follows the theoretical prediction given by the asymptotic and numerical models, as previously shown.
\begin{figure}[h]
  \begin{center}
      \includegraphics[width= 15 cm]{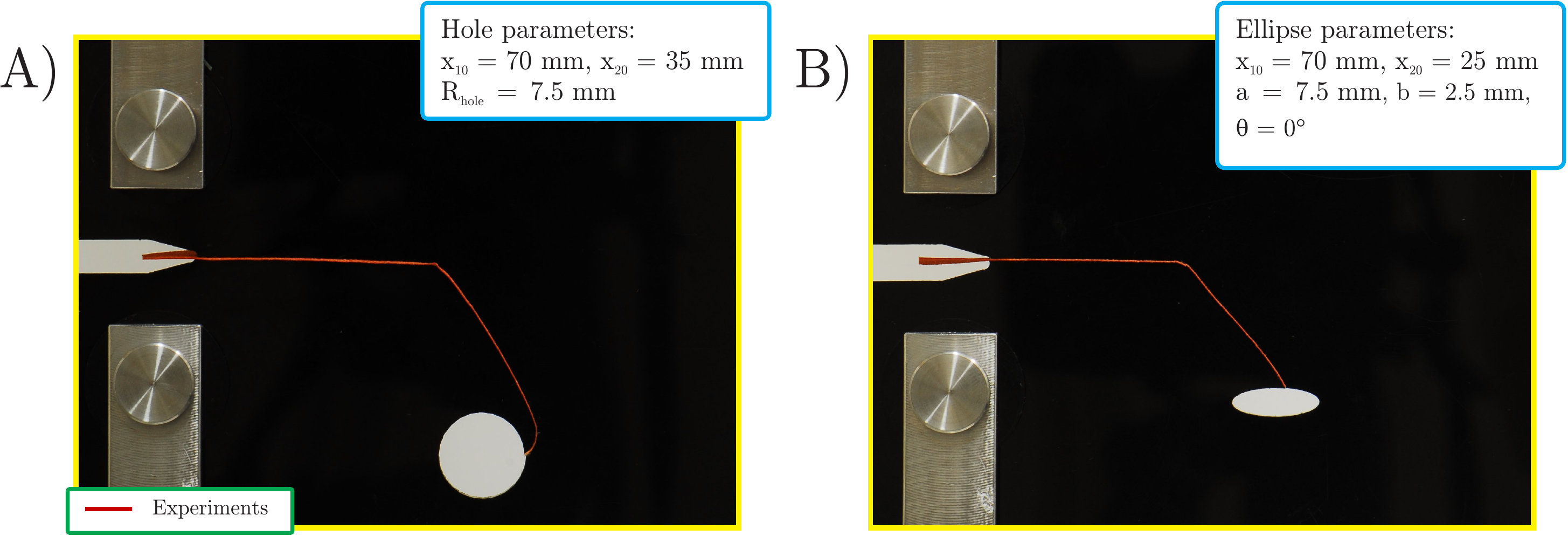}
\caption{\footnotesize {Two examples of dynamic kinking during fracture propagation. The crack trajectory is highlighted by means of a solid/red colored strip inserted between the two crack surfaces in the post mortem samples. }} \lb{Dynamic_VS_Static}
  \end{center}
\end{figure}
Fig. \ref{Dynamic_VS_Static} shows kinking instability in the case of dynamic crack growth as related to the roughness of the crack path shown in Fig. \ref{surf} (B), to be compared
to the smoothness of the paths found in all cases of quasi-static propagation (see for example Fig. \ref{surf} A).
\begin{figure}[h]
  \begin{center}
      \includegraphics[width= 15 cm]{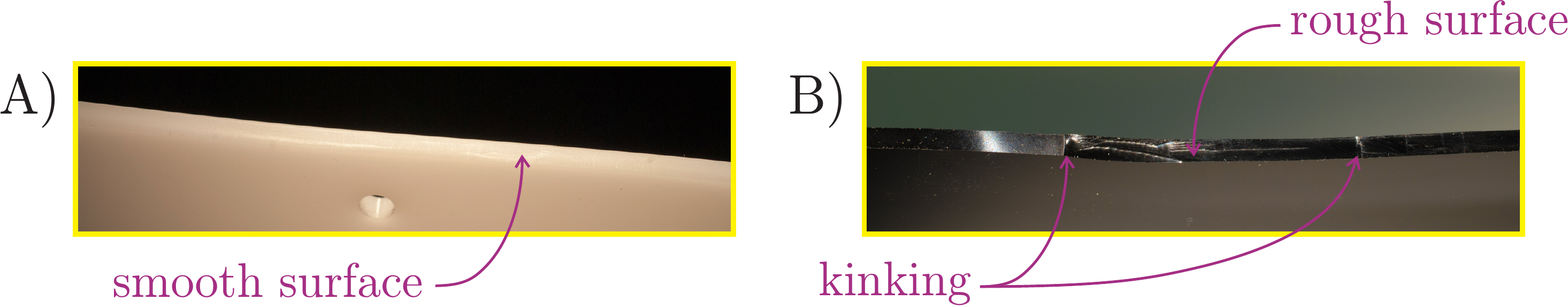}
\caption{\footnotesize { Comparison between the fracture surfaces during quasi-static (A) and dynamic (B) propagation. In a quasi-static regime (A) the crack path is smooth, while a dynamic regime (B) crack growth is characterized by roughness of the fracture surface and kinking.}} \lb{surf}
  \end{center}
\end{figure}

\section{Conclusions}

An experimental and computational (using adaptive finite elements with an X-FEM algorithm and moving boundary associated with an advancing crack) verification has been presented for an asymptotic model, developed to predict the path of a crack propagating in a brittle material.
In the quasi-static regime, the predictions of the asymptotic model are fairly close to experiments.
When the size of the voids becomes large compared to the distance between the unperturbed crack trajectory and the center of the voids, the approximation error of the asymptotic model also increases.
Moreover, when the crack propagation reaches a critical dynamic regime, kinking of the crack path occurs, and consequently, the crack may be arrested as a result of hitting one of the voids.
Figs. \ref{palla} and \ref{Dynamic_VS_Static} illustrate these special cases when the crack deviates from the asymptotic model. The experiments have demonstrated that, when the underlying hypotheses remain valid,  the asymptotic model gives excellent results for a wide range of loading rate conditions, as shown in Figs. \ref{Exp_holes}, \ref{Exp_ellipses_1} and \ref{Exp_ellipses2}.
The importance of this analysis lies in the possibility of accurately predicting the failure patterns of brittle materials, such as ceramics, with inhomogeneities in the form of small inclusions or voids.


\vspace*{5 mm} \noindent {\sl Acknowledgments }

This work was motivated by the recent studies of fracture in ceramics in the framework of the European FP7 - INTERCER-2
project (PIAP-GA-2011-286110-INTERCER2). From this grant D.M., A.B.M., and N.V.M. acknowledge financial support.
D.B. acknowledges support from the ERC Advanced Grant  \lq Instabilities and nonlocal multiscale modelling
of materials' (ERC-2013-ADG-340561-INSTABILITIES)

\vspace*{5mm} \noindent \vspace*{10mm}

 { \singlespace
}

\end{document}